\documentclass[12pt]{iopart}
\usepackage{graphicx}
\usepackage{cite}
\expandafter\let\csname equation*\endcsname\relax
\expandafter\let\csname endequation*\endcsname\relax
\usepackage{amsmath}
\DeclareMathOperator*{\argmax}{arg\,max}
\usepackage{iopams}
\usepackage{url}
\usepackage{bm}
\usepackage{enumitem}
\usepackage{array}

\usepackage{etoolbox}

\makeatletter
\newcommand{\mainmatter}{%
	\setcounter{footnote}{0}%
	\patchcmd{\@makefntext}{\fnsymbol}{\arabic}{}{}%
	\patchcmd{\@thefnmark}{\fnsymbol}{\arabic}{}{}%
	\def\@makefnmark{\textsuperscript{\arabic{footnote}}}%
}
\makeatother

\begin{document}

\title[]{On the convergence of phase space distributions to microcanonical equilibrium: dynamical isometry and generalized coarse-graining}

\author{Casey O. Barkan}

\address{Department of Physics and Astronomy, University of California, Los Angeles}
\ead{barkanc@ucla.edu}
\vspace{10pt}
\begin{indented}
\item[]August 2024
\end{indented}

\begin{abstract} 
This work explores the manner in which classical phase space distribution functions converge to the microcanonical distribution. We first prove a theorem about the lack of convergence, then define a generalization of the coarse-graining procedure that leads to convergence. We prove that the time evolution of phase space distributions is an isometry for a broad class of statistical distance metrics, implying that ensembles do not get any closer to (or farther from) equilibrium, according to these metrics. This extends the known result that strong convergence of phase space distributions to the microcanonical distribution does not occur. However, it has long been known that weak convergence can occur, such that coarse-grained distributions---defined by partitioning phase space into a finite number of cells---converge pointwise to the microcanonical distribution. We define a generalization of coarse-graining that removes the need for partitioning phase space into cells. We prove that our generalized coarse-grained distribution converges pointwise to the microcanonical distribution if the dynamics are strong mixing. As an example, we study an ensemble of triangular billiard systems.

\end{abstract}

% Keywords
\vspace{2pc}

\mainmatter

\section{Introduction}
The dynamics of classical phase space distribution functions have been widely studied with the aim of providing a rigorous description of a system’s approach to thermal equilibrium \cite{gibbs1902elementary, Ehrenfest1912,mackey1989dynamic,dougherty1994foundations,kozlov2007fine,kumivcak1998dynamics,castagnino1998dynamics,ruelle1999smooth,lewis1967unifying,Krylov1979,courbage1990markov,kozlov2003evolution,kozlov2003weak,piftankin2010coarse,trushechkin2010irreversibility,piftankin2008gibbs,kozlov2015coarsening,kalogeropoulos2022coarse,dougherty2003lyapunov}. 
Phase space distributions describe ensembles of systems obeying Hamilton’s equations, and the dynamics of these distributions are described by Liouville’s equation\cite{reichl2016modern}. A central question is: Under what conditions does the phase space distribution approach the microcanonical distribution $\rho^\text{MC}$ as time $t\to\infty$? It has long been known that for Hamiltonians with a property called strong mixing, phase space distributions approach $\rho^\text{MC}$ with \textit{weak} convergence on each constant-energy hypersurface of phase space, but stronger forms of convergence do not occur \cite{mackey1989dynamic, dougherty1994foundations, kumivcak1998dynamics, kozlov2007fine, castagnino1998dynamics}. In fact, it has been shown that for certain metrics of statistical distance, the distance between any two distributions is constant in time \cite{kumivcak1998dynamics, daffertshofer2002classical,plastino2004liouville,yamano2008classical,yamano2013bruijn}. In this paper, we prove a theorem that generalizes these results to (i) a broader class of statistical distance metrics, and (ii) distributions that are not continuous on phase space, which is necessary for defining the statistical distance to $\rho^\text{MC}$ ($\rho^\text{MC}$ is discontinuous because it is nonzero on a constant-energy hypersurface and zero elsewhere).
%In this paper, we extend these results by proving that the distance is constant in time for a broad class of metrics for statistical distance.

A closely related question is how (and whether) the entropy converges to its maximal value, the microcanonical entropy. It is well known that the Gibbs entropy of any distribution is constant in time under the dynamics of Liouville’s equation \cite{gibbs1902elementary,Ehrenfest1912,mackey1989dynamic,dougherty1994foundations,kumivcak1998dynamics,ruelle1999smooth,kozlov2007fine,castagnino1998dynamics}. This fact  may appear paradoxical in light of the 2nd law of thermodynamics, and Gibbs proposed to resolve this paradox by coarse-graining phase space, i.e. partitioning phase space into a finite number of cells (Chapter XII of \cite{gibbs1902elementary}). Strong mixing guarantees that the coarse-grained entropy converges to the microcanonical entropy as $t\to\infty$ (and as $t\to-\infty$, due to time reversal symmetry) \cite{dougherty1994foundations,mackey1989dynamic,kozlov2007fine, safranek2020classical}. However, the partition of phase space used for the coarse-graining is chosen arbitrarily and is not well-motivated physically; this contrasts with weak convergence of phase space distributions to $\rho^\text{MC}$, which does not involve any partitioning of phase space. In this work, we define a generalization of coarse-graining which removes the need for partitioning phase space, and we prove that this generalized coarse-grained distribution converges pointwise to $\rho^\text{MC}$ for strong mixing systems. Our generalized coarse-grained distribution has similarities to the maximum entropy distribution introduced by E.T. Jaynes \cite{jaynes1957information}, though there are key differences in their dynamical behaviors, as we discuss.

In much of the literature on the dynamics of phase space distributions, a primary motivation has been to provide a rigorous understanding of thermalization and the 2nd law of thermodynamics \cite{gibbs1902elementary,Ehrenfest1912,mackey1989dynamic,dougherty1994foundations,Krylov1979,castagnino1998dynamics,kumivcak1998dynamics,safranek2020classical,kozlov2007fine,trushechkin2010irreversibility,piftankin2008gibbs,piftankin2010coarse}. Yet, compelling arguments have been made that the dynamics of phase space distributions does not provide a complete explanation of thermalization \cite{goldstein2001boltzmann,goldstein2019individualist,goldstein2019nonequilibrium,bricmont2001bayes,albert2003time,tumulka2019}. The reasoning is that phase space distributions describe \textit{ensembles} of identical and non-interacting systems, whereas it is an experimental fact that \textit{individual} isolated macroscopic systems reach thermal equilibrium; hence, it should be possible to explain thermalization in terms of the dynamics of individual systems, not ensembles. These differing perspectives have been termed the ``Ensemblist" view and the ``Individualist" view \cite{goldstein2019individualist,tumulka2019}. 
We discuss the physical scenarios in which the dynamics and weak convergence of phase space distributions are relevant, contending that these dynamics are relevant to particular computational methods and experiments, but that they do not provide a complete account of the thermalization of macroscopic systems.

%For the purposes of this paper, we adopt an Ensemblist definition of thermalization: 

\section{Background definitions and concepts}

\subsection{Liouville's equation}
Consider a classical mechanical system with phase space $\mathcal{P}$ and Hamiltonian $H(\vec x)$. The notation $\vec x = (q_1,...,q_N,p_1,...p_N)$ denotes a point in phase space, where $q_i$ and $p_i$ are canonincal position and momentum coordinates. Let $\phi_t:\mathcal{P}\to \mathcal{P}$ denote the time evolution map that evolves the system forward by time $t$. In other words, $\vec x(t) = \phi_t(\vec x_0)$ where $\vec x(t)$ is a trajectory of Hamilton's equations with initial condition $\vec x_0$.

An ensemble of such systems is described by a phase space distribution function $\rho_t(\vec x)$, which is a probability density function on $\mathcal{P}$ normalized so that $\int_\mathcal{P} \rho_t(\vec x)d\vec x=1$. The dynamics of $\rho_t(\vec x)$ obey  \cite{dougherty1994foundations}
\begin{equation}\label{my_Liouville}
    \rho_t(\vec x)=\rho_0(\phi_{-t}(\vec x))
\end{equation}
where $\rho_0(\vec x)$ is the distribution at $t=0$. We call this equation \textbf{\textit{Liouville's equation}} \footnote{It is more common to call Eq.~\ref{Liouville} Liouville's Equation, but Eq.~\ref{my_Liouville} is more general.}. Intuitively, Liouville's equation states that probability density remains constant in time along trajectories of Hamilton's equations. If $\rho_t(\vec x)$ is differentiable, then \cite{reichl2016modern}
\begin{equation}\label{Liouville}
	\frac{\partial}{\partial t} \rho_t(\vec x) = -\vec v(\vec x)\cdot\nabla\rho_t(\vec x)
\end{equation}
where $\vec v(\vec x)$ is the phase space ``velocity" given by Hamilton's equations: $v_i(\vec x) = \partial H(\vec x)/\partial p_i$ for $i\in\{1,...,N\}$ and $v_i(\vec x) = -\partial H(\vec x)/\partial q_i$ for $i\in\{N+1,...,2N\}$.
%\begin{equation}
%	\vec v(q_1,...,q_N,p_1,...,p_N) = \left( \frac{\partial H}{\partial p_1},...,\frac{\partial H}{\partial p_N},-\frac{\partial H}{\partial q_1},...,-\frac{\partial H}{\partial q_N} \right)
%\end{equation}

\subsection{Invariant sets and invariant measures}\label{invariant}

We will consider distribution functions with support not only on the full phase space $\mathcal{P}$, but also on subsets $P\subseteq\mathcal{P}$. An important example is the microcanonical distribution, which has support on a constant-energy hypersurface,
\begin{equation}
	\Sigma_E=\{\vec x\in\mathcal{P}:H(\vec x)=E\}
\end{equation}
We will consider only invariant sets ($P$ is an \textbf{\textit{invariant set}} if $\phi_t(P)= P$ for all $t$ \footnote{For general dynamical systems, $P$ is an invariant set if $\phi_{t}(P)\subseteq P$ for $t>0$. For Hamiltonian systems, this is equivalent to $\phi_t(P)=P$.}), which ensures that if a distribution has support in $P$ at $t=0$, its support will remain in $P$ for all $t$.

Integrals over $\mathcal{P}$ utilize the Lebesgue measure  $d\vec x\equiv d\vec qd\vec p$, but integrals over subsets $P\subset\mathcal{P}$ for which $\int_Pd\vec x=0$ (such as $\Sigma_E$) require an appropriate projection of the Lebesgue measure onto $P$. For $\Sigma_E$, this measure is $d\mu=d\Sigma/||\nabla H(\vec x)||$ where $d\Sigma$ is the Lebesgue surface measure \cite{khinchin1949mathematical,arnold1968ergodic}. For any $P$, the appropriate measure is an \textbf{\textit{invariant measure}}, meaning that $\mu(A)=\mu(\phi_t(A))$ for all $t$ and for any measureable $A\subseteq P$, where $\mu(A)\equiv \int_Ad\mu$ \footnote{Liouville's theorem implies that the Lebesgue measure is invariant on $\mathcal{P}$, and the projection of the Lebesgue measure onto any subset $P$ must also be invariant.}. In the following, we use $d\mu$ to denote the invariant measure for whichever $P$ is under consideration.

\subsection{Strong mixing}\label{mixing}
%Let $\Sigma_E$ denote a hypersurface of $\mathcal P$ with constant energy $E$, i.e. $\Sigma_E=\{ \vec x\in\mathcal{P}: H(\vec x)=E \}$.
The dynamics on a constant-energy hypersurface $\Sigma_E$ are \textbf{\textit{strong mixing}} if $\lim_{t\to\infty} \frac{\mu(\phi_t(A) \cap B)}{\mu(A)} = \frac{\mu(B)}{\mu(\Sigma_E)}$ for any $A,B\subseteq\Sigma_E$ with $\mu(A)>0$ \cite{arnold1968ergodic, nicol2009}. To understand this definition intuitively, consider a system with initial condition $\vec x_0$ drawn uniformly at random from $A$. Then, in the $t\to\infty$ limit, the probability that $x(t)$ is in some region $B$ is merely the fraction of $\Sigma_E$ covered by $B$. In other words, the system ``forgets" its initial condition. The related concept of weak mixing \cite{nicol2009} arises if one considers time-averaging, but we will not utilize this concept here.

\subsection{Strong convergence}

Consider a space of functions $\mathcal{F}$ with a distance metric $\text{dist}[f,g]\in\mathbb{R}_{\geq0}$ for $f,g\in\mathcal{F}$. A family of functions $f_t(\vec x)$ (parametrized by $t$) approaches a limit $f_{Eq.}(\vec x)$ as $t\to\infty$ with \textbf{\textit{strong convergence}} if $\text{dist}[f_t,f_{Eq.}]\to0$ as $t\to\infty$. In section \ref{strong}, we will consider a broad class of distance metrics on the space of phase space distributions.

\subsection{Weak convergence}

Weak convergence of a distribution $\rho_t(\vec x)$ to an equilibrium distribution means that ensemble averages of observables converge to their equilibrium values. More precisely, consider a distribution $\rho_t(\vec x)$ on $P\subseteq \mathcal{P}$. $\rho_t(\vec x)$ approaches $\rho_{Eq.}(\vec x)$ with \textbf{\textit{weak convergence}} if, for all square integrable functions $a(\vec x):P\to\mathbb{R}$ (called ``observables"), $\langle a\rangle_{\rho_t} \to \langle a\rangle_{\rho_\text{Eq.}}$ as $t\to\infty$. The notation $\langle a\rangle_{\eta}$ denotes the ensemble average of $a(\vec x)$ according to a distribution $\eta(\vec x)$,
\begin{equation}\label{ens_avg}
	\langle a\rangle_\eta = \int_{P} a(\vec x)\eta(\vec x) d\mu
\end{equation}
An example of a function that converges to a limit weakly, but not strongly, is discussed in the Supplementary Material.

\subsection{Equivalence of strong mixing and weak convergence to $\rho^\text{MC}$}\label{weak_and_strong}
Strong mixing guarantees that phase space distributions on a constant-energy hypersurface weakly converge to the microcanonical distribution $\rho^\text{MC}$, given by 
\begin{equation}
	\rho^\text{MC}=\frac{1}{\mu(\Sigma_E)}
\end{equation}
with $d\mu=d\Sigma/||\nabla H(\vec x)||$. This is the uniform distribution on $\Sigma_E$.

More precisly, every square integrable (and normalized) distribution $\rho_t(\vec x)$ on a hypersurface $\Sigma_E$ converges weakly to $\rho^\text{MC}$ if and only if the dynamics on $\Sigma_E$ are strong mixing (see Theorem 9.8 in \cite{arnold1968ergodic}). In fact, some authors take weak convergence to $\rho^\text{MC}$ as the definition of strong mixing \cite{reichl2016modern}.

\section{The isometry of phase space distribution dynamics} \label{strong}

Are there conditions under which $\rho_t(\vec x)$ approaches an equilibrium distribution as $t\to\infty$? The answer to this question depends on how one defines ``approaches". In fact, for several metrics of statistical distance, it has been shown that the distance between two distributions is contant in time \cite{kumivcak1998dynamics, daffertshofer2002classical, plastino2004liouville,yamano2008classical,yamano2013bruijn}. Hence, according to these metrics, time evolution is an isometry and strong convergence to an equilibrium distribution cannot occur. This has been shown for the $L^2$ distance \cite{kumivcak1998dynamics}, the Kullback-Leibler divergence \cite{daffertshofer2002classical} (and a generalization thereof \cite{plastino2004liouville}), the Renyi-$\alpha$ divergence \cite{yamano2008classical}, and the Hellinger distance \cite{yamano2013bruijn}. We generalize these results by proving that the distance is constant for a broad class of metrics, which includes all of the above listed metrics as special cases. Additionally, our proof generalizes these previous results in that it does not assume differentiability or continuity of the distributions. This is essential when considering the distance to microcanonical equilibrium, because the microcanonical distribution is not continuous on phase space $\mathcal{P}$.

Consider the class of distance metrics \footnote{This class includes functionals that are not formally metrics (i.e. that don't satisfy $\text{dist}[f,g]=\text{dist}[g,f]$ or $\text{dist}[f,f]=0$). Our results hold for all functionals of the form of Eqs.~\ref{dist} and \ref{I}, regardless of whether they satisfy the formal criteria for a metric.} of the form
\begin{equation}\label{dist}
	\text{dist}[f(\vec x),g(\vec x)] = d(I[f(\vec x),g(\vec x)])
\end{equation}
where $d:\mathbb{R}\to\mathbb{R}$ and
\begin{equation}\label{I}
	I[f(\vec x),g(\vec x)] = \int_P F(f(\vec x),g(\vec x)) d\mu
\end{equation}
where $F:\mathbb{R}^2\to\mathbb{R}_{\geq0}$. Table 1 lists several standard distance metrics of this form. Theorem 1 states our first result: The distance between any two distributions is constant in time according to any metric defined by Eqs.~\ref{dist} and \ref{I}. In other words, time evolution of phase space distributions is an isometry. Theorem 1 is valid for any Hamiltonian with any number of degrees of freedom, and it implies that $\rho_t(\vec x)$ does not strongly converge to any limit.
\medskip \\
\noindent\textbf{Theorem 1: } Let $\rho_t(\vec x)$ and $\eta_t(\vec x)$ be two phase space distributions obeying Liouville's equation and define $\text{dist}[\rho_t(\vec x),\eta_t(\vec x)]$ according to Eqs.~\ref{dist} and \ref{I} for any $d:\mathbb{R}\to\mathbb{R}$, any Lebesgue integrable $F:\mathbb{R}\to\mathbb{R}_{\geq0}$, and any invariant $P\subseteq\mathcal{P}$ with $d\mu$ the corresponding invariant measure. Then, $\text{dist}[\rho_t(\vec x),\eta_t(\vec x)]$ is independent of time $t$.
\smallskip \\
\noindent\textbf{Proof: } See Appendix A.
\smallskip \\

\begin{table}
	\caption{Standard metrics of distance between probability distributions. The distance $\text{dist}[f(\vec x),g(\vec x)] = d(I[f(\vec x),g(\vec x)])$ where $I=\int_P F(f(\vec x),g(\vec x))d\mu$ and the functions $F$ and $d$ are listed in the table. The last three rows are common entropies to which Theorem 1 can also be applied.}
	\footnotesize
	\begin{tabular}{lcc}
		\br
		Metrics, divergences, and entropies & $F(f,g)$ & $d(I)$ \\
		\mr
		Total variation distance \cite{van2014renyi} & $|f-g|$ & $I/2$ \\
		$L^p$ distance ($p\geq1$) \cite{EOMLpSpaces} & $|f-g|^p$ & $I^{1/p}$ \\
		Hellinger distance \cite{van2014renyi} & $(\sqrt{f}-\sqrt{g})^2$ & $I/2$ \\
		Cross entropy$^{a}$ \cite{cover1999elements} & $f\log g$ & $-I$ \\
		Kullback-Leibler divergence$^{a}$ \cite{cover1999elements} & $f \log(f/g)$ & $I$ \\
		Renyi $\alpha$-divergence$^a$ \cite{van2014renyi} & $f^\alpha/g^{\alpha-1}$ & $\frac{1}{\alpha-1}\log(I)$ \\
		Gibbs (Shannon) entropy \cite{gibbs1902elementary} & $f\log(f)$ & $- I$ \\
		Renyi $\alpha$-entropy \cite{van2014renyi} & $f^\alpha$ & $\frac{1}{\alpha-1}\log(I)$ \\
		Tsallis $\alpha$-entropy \cite{tsallis1988possible} & $f^\alpha$ & $\frac{1}{\alpha-1}(1-I)$ \\
		\br
	\end{tabular}\\
	$^{a}$ These are defined only when the support of $f$ is a subset of the support of $g$, and $F(0,0)$ is defined to be $0$.
\end{table}
\normalsize

Theorem 1 can also be used to prove that various definitions of entropy are contstant in time for $\rho_t(\vec x)$ obeying Liouville's equation. For example, it is well-known that the Gibbs entropy is constant in time \cite{gibbs1902elementary, Ehrenfest1912,mackey1989dynamic, dougherty1994foundations,kumivcak1998dynamics,ruelle1999smooth,kozlov2007fine,castagnino1998dynamics}, where Gibbs entropy is defined as
\begin{equation}\label{S}
	S[\rho_t(\vec x)] = -\int_P \rho_t(\vec x)\log(\rho_t(\vec x))d\mu
\end{equation}
where we use $k_B=1$. To establish the connection with Theorem 1, let $F(f,g)=f\log(f)$ and $d(I)=-I$. Then, Theorem 1 implies that $S[\rho_t(\vec x)]$ is independent of $t$. Similarly, Theorem 1 implies that the Renyi $\alpha$-entropy and Tsallis $\alpha$-entropy are constant in time (see Table 1 for the $F(f,g)$ and $d(I)$ for these entropies).

Despite the lack of strong convergence, weak convergence to $\rho^\text{MC}$ occurs if the dynamics are strong mixing (see section \ref{weak_and_strong}). A simple example showing how a function can converge to a limit weakly, but not strongly, is given in the Supplementary Material.

When measurement precision and/or computational power is limited, weak convergence is the appropriate notion of convergence. For example, consider numerically solving Eq.~\ref{Liouville} where space is discretized as a grid. For strong mixing dynamics, gradients of $\rho_t(\vec x)$ diverge as $t\to\infty$, so the numerical algorithm cannot compute $\rho_t(\vec x)$ to an accuracy that will preserve the distance $\text{dist}[\rho_t(\vec x),\rho^\text{MC}]$ for arbitrarily long times. Hence, Theorem 1, although formally true, becomes irrelevant due to the limitation in numerical precision. Similarly, if experimental measurement precision is limited such that only ensemble averages of certain observables can be measured, then strong mixing dynamics imply that $\rho_t(\vec x)$ will become experimentally indistinguishable from $\rho^\text{MC}$, despite Theorem 1. These considerations motivate imposing a coarse-graining on $\rho_t(\vec x)$, as discussed in the following section.

\section{Generalization of coarse-graining}

Coarse-graining the distribution $\rho_t(\vec x)$ is a widely proposed method of accounting for imperfect measurement precision \cite{gibbs1902elementary,Ehrenfest1912,mackey1989dynamic,kozlov2007fine,safranek2020classical,piftankin2008gibbs,piftankin2010coarse,kozlov2015coarsening}. Gibbs first proposed coarse-graining to resolve the paradox that $S[\rho_t(\vec x)]$ is constant in time (Chapter XII of \cite{gibbs1902elementary}). Coarse-graining involves a partition of phase space into ``cells", and $\rho_t(\vec x)$ is averaged over each cell to produce a coarse-grained distribution $\tilde \rho_t(\vec x)$; this procedure is described in more detail in section \ref{standard}. On a constant-energy hypersurface $\Sigma_E$, $\tilde\rho_t(\vec x)$ converges \textit{pointwise} to the microcanonical distribution $\rho^\text{MC}$ when the dynamics are strong mixing \cite{kozlov2007fine,safranek2020classical}. Consequently, the coarse-grained entropy $S[\tilde\rho_t(\vec x)]$ increases to its maximal value of $S[\rho^\text{MC}]$ as $t\to\infty$ (and as $t\to-\infty$), resolving the apparent paradox.
%Here and in the following, $S[\rho]$ is defined by Eq.~\ref{S} where $p=\Sigma_E$ and $d\mu=d\Sigma/||\nabla H(\vec x)||$.

There are two similar, yet distinct, results for strong mixing systems that are important to compare: (i) $\rho_t(\vec x)$ converges weakly to $\rho^\text{MC}$, and (ii) $\tilde\rho_t(\vec x)$ converges pointwise to $\rho^\text{MC}$. Result (i) has the advantage that it does not require any partitioning of phase space, but it has the disadvantage that it fails to account for the 2nd law because $S[\rho_t(\vec x)]$ is constant in time despite weak convergence. Result (ii) has the advantage that $S[\tilde\rho_t(\vec x)]\to S[\rho^\text{MC}]$, but it has the disadvantage that it requires partitioning phase space, which is not physically well-motivated.

We propose a generalization of coarse-graining that keeps the advantages of the above results (i) and (ii) while remedying their disadvantages. Akin to weak convergence, our generalization is defined in terms of ensemble averages of observables, putting generalized coarse-graining and weak convergence on the same footing.
The generalized coarse-grained distribution, denoted $\rho^\text{G}_t(\vec x)$, is defined by a maximum-entropy optimization problem, similar to E.T. Jaynes' maximum-entropy distribution \cite{jaynes1957information}. However, there is a key difference between $\rho^\text{G}_t(\vec x)$ and Jaynes' distribution, namely, $\rho^\text{G}_t(\vec x)$ converges to the microcanonical distribution for strong mixing dynamics (Theorem 2 below), whereas Jaynes' distribution does not converge to either the canonical nor microcanonical distribution except under restrictive assumptions (see Appendix B). In sections \ref{standard}, \ref{generalized}, and \ref{sim_section} below, we first introduce standard coarse-graining, then define our generalization of coarse-graining, and lastly we compare the two forms of coarse-graining with an example.

In all that follows, we only consider distributions on a hypersurface $\Sigma_E$. Physically, we assume that the energy $E$ can be determined experimentally (or in simulations, set numerically). Consequently, $d\mu=d\Sigma/||\nabla H(\vec x)||$, so entropy is defined as
\begin{equation}\label{surfS}
	S[\rho_t(\vec x)] = -\int_{\Sigma_E} \rho_t(\vec x)\log \rho_t(\vec x) \frac{d\Sigma}{||\nabla H(\vec x)||}
\end{equation}

\subsection{Standard coarse-graining}\label{standard}

The standard definition of coarse-graining involves a partition of $\Sigma_E$ into $M$ cells $\Gamma_i$, $i=1,...,M$, satisfying $\cup_{i=1}^M\Gamma_i=\Sigma_E$ and $\Gamma_i\cap\Gamma_j=\emptyset$ for $i\neq j$. The partition is intended to capture the fact that real-world measurements are imperfect, the idea being that each cell is a collection of points that are indistinguishable to the imperfect measurement device. However, in section \ref{generalized} we show that this is just a special case of a more natural and general approach for describing imperfect measurements.

The standard coarse-grained distribution is defined as
\begin{equation}\label{SCG}
	\tilde\rho_t(\vec x)=\frac{1}{\mu(\Gamma_{i(\vec x)})}\int_{\Gamma_{i(\vec x)}}\rho_t(\vec x')d\mu
\end{equation}
where $i(\vec x)$ is the index of the cell containing the point $\vec x$, and $\rho_t(\vec x)$ is the true distribution which obeys Liouville's equation and which determines experimental observations.

Equivalently, $\tilde\rho_t(\vec x)$ can be defined by a maximum entropy optimization problem. Let $1_{\Gamma_i}(\vec x)$ denote the indicator function of the set $\Gamma_i$, defined as $1_{\Gamma_i}(\vec x)=1$ if $\vec x\in\Gamma_i$ and $1_{\Gamma_i}(\vec x)=0$ otherwise. Then,
\begin{equation}\label{maxS}
\begin{split}
	\tilde\rho_t(\vec x) = \argmax_{\eta(\vec x)\in L^2}\; S[\eta(\vec x)] \quad \text{subject to}& \quad \int_{\Sigma_E} \eta(\vec x)d\mu = 1 \\
	\text{and}& \quad \langle 1_{\Gamma_i} \rangle_{\eta} = \langle 1_{\Gamma_i} \rangle_{\rho_t} \text{ for } i=1,...,M
\end{split}
\end{equation}
where the ensemble averages $\langle 1_{\Gamma_i} \rangle_{\eta}$ and $\langle 1_{\Gamma_i} \rangle_{\rho_t}$ are defined according to Eq.~\ref{ens_avg} with $P=\Sigma_E$, and $L^2$ denotes the space of square integrable functions on $\Sigma_E$. Note that
$\langle 1_{\Gamma_i} \rangle_{\rho_t}$ are experimentally observable ensemble averages and are treated as known values when solving the optimization. The optimization problem is to be solved at each time $t$ to obtain $\tilde\rho_t(\vec x)$. In Appendix C.1 it is shown that the two definitions, Eq.~\ref{SCG} and Eq.~\ref{maxS}, are equivalent. Eq.~\ref{maxS} clarifies the motivation for coarse-graining: $\tilde\rho_t(\vec x)$ is the highest entropy distribution consistent with the information available via measurements of ensemble averages.

%Eq.~\ref{maxS} makes clear that for strong mixing systems, $\tilde\rho(\vec x)\to\rho^\text{MC}$ pointwise as $t\to\infty$. This is because strong mixing implies weak convergence, hence $\langle 1_{\Gamma_i} \rangle_{\tilde\rho_t} = \langle 1_{\Gamma_i} \rangle_{\rho_t} \to  \langle 1_{\Gamma_i} \rangle_{\rho^\text{MC}}$. Consequently, the coarse-grained entropy $S[\tilde\rho(\vec x)]\to S[\rho^\text{MC}]$ as $t\to\infty$. See Theorem 2 below for a proof of this statement in a more general form.

\subsection{Generalized coarse-graining}\label{generalized}

Eq.~\ref{maxS} suggests a natural generalization of coarse-graining. Rather than assuming that measurement capability is characterized by indicator functions $1_{\Gamma_i}(\vec x)$, we allow for measurement capabilities characterized by any list of observables $a_i(\vec x)$, $i=1,...,M$. Accordingly, we define the generalized coarse-grained distribution as the function $\rho^\text{G}_t(\vec x):\Sigma_E\to\mathbb{R}_{\geq 0}$ that solves the following optimization problem:
\begin{equation}\label{rhoG}
	\begin{split}
		\rho^\text{G}_t(\vec x) = \argmax_{\eta(\vec x)\in L^2}\; S[\eta(\vec x)] \quad \text{subject to}& \quad \int_{\Sigma_E} \eta(\vec x)d\mu = 1 \\
		\text{and}& \quad \langle a_i \rangle_{\eta} = \langle a_i \rangle_{\rho_t} \text{ for } i=1,...,M
	\end{split}
\end{equation}
where, as before, $\langle a_i \rangle_{\eta}$ and $\langle a_i \rangle_{\rho_t}$ are defined according to Eq.~\ref{ens_avg} with $P=\Sigma_E$, and $L^2$ is the space of square integrable functions on $\Sigma_E$. $\langle a_i \rangle_{\rho_t}$ are experimentally observed so are treated as known values, and the optimization problem is to be solved at each time $t$ to obtain $\rho^\text{G}_t(\vec x)$. Appendix B compares and contrasts $\rho^\text{G}_t(\vec x)$ with the maximum entropy distribution introduced by E.T. Jaynes. The following theorem captures the key behaviors of $\rho^\text{G}_t(\vec x)$.

\medskip
\noindent\textbf{Theorem 2: } Let $\rho_t(\vec x)$ be a normalized and square integrable distribution on $\Sigma_E$ obeying Liouville's equation for a strong mixing system. Define $\rho^\text{G}_t(\vec x)$ according to Eq.~\ref{rhoG}. Then,
\begin{enumerate}
	\item $\lim_{t\to\infty}\rho^\text{G}_t(\vec x)=\rho^\text{MC}$ for all $\vec x\in\Sigma_E$, i.e. pointwise convergence.
	\item $\lim_{t\to\infty}S[\rho^\text{G}_t(\vec x)] = S[\rho^\text{MC}]$.
\end{enumerate}
\smallskip
\noindent\textbf{Proof: } See Appendix D.
\smallskip \\

A fundamental aspect of the coarse-graining procedure is that the distribution $\rho^\text{G}_t$ and entropy $S[\rho^\text{G}_t]$ depend upon the information available to the observer. Hence, different observers may assign different entropies to the same ensemble. In particular, adding additional observables to Eq.~\ref{rhoG} decreases the entropy of $\rho^\text{G}_t$ (or leaves the entropy unchanged). This reflects the fact that, with more observables, it is easier to perceive discrepancies between $\rho^\text{G}_t$ and $\rho^\text{MC}$; in other words, additional observables reduce one's ignorance of the system's state, thereby reducing entropy. Importantly, although different observers may disagree about the entropy at finite $t$, all observers agree that $S[\rho^\text{G}_t]$ converges to $S[\rho^\text{MC}]$ in the $t\to\infty$ limit, if the dynamics are strong mixing.

The optimization problem that defines $\rho^\text{G}_t$ can be solved using the method of the Lagrange dual function \cite{boyd2004convex}; see Appendix C for details. The result is
\begin{equation}\label{rhoG_solved}
	\rho^\text{G}_t(\vec x) = \exp\left(-1-\lambda_{0}^*(t) - \sum_{i=1}^M \lambda_{i}^*(t) a_i(\vec x)\right)
\end{equation}
where, at each $t$, the values $\lambda_{i}^*(t)$ are defined as those that minimize the following function:
\begin{equation}\label{g}
	g_t(\lambda_0,...,\lambda_M) = \int_{\Sigma_E} \exp\left( -1-\lambda_{0}-\sum_{i=1}^M\lambda_ia_i(\vec x) \right)d\mu + \lambda_0 + \sum_{i=1}^M \lambda_i\langle a_i\rangle_{\rho_t}
\end{equation}
The values $\lambda_{i}^*(t)$ are the Lagrange multipliers and the function $g_t(\lambda_0,...,\lambda_M)$ is the Lagrange dual function corresponding to Eq.~\ref{rhoG}. The Lagrange dual function is guaranteed to be convex with a global minimum and no other local minima \cite{boyd2004convex}. As a result, a gradient descent algorithm is guaranteed to converge to the unique minimum. Note that Eqs.~\ref{rhoG_solved} and $\ref{g}$ reduce the original infinite dimensional constrained optimization problem (Eq.~\ref{rhoG}) to a finite dimensional unconstrained optimization problem for which standard algorithms are guaranteed to converge.

The entropy of $\rho^\text{G}_t(\vec x)$ can be written in a simple form in terms of the Lagrange multipliers: $S[\rho^\text{G}_t(\vec x)] = 1 + \lambda_{0}^*(t) + \sum_{i=1}^M \lambda_{i}^*(t)\langle a_i\rangle_{\rho_t}$. An analogous expression was found by Jaynes for the entropy of his maximum entropy distribution \cite{jaynes1957information}.

\subsection{Example: An ensemble of triangular billiards}\label{sim_section}

\begin{figure}
    \centering
    \includegraphics[width=1\linewidth]{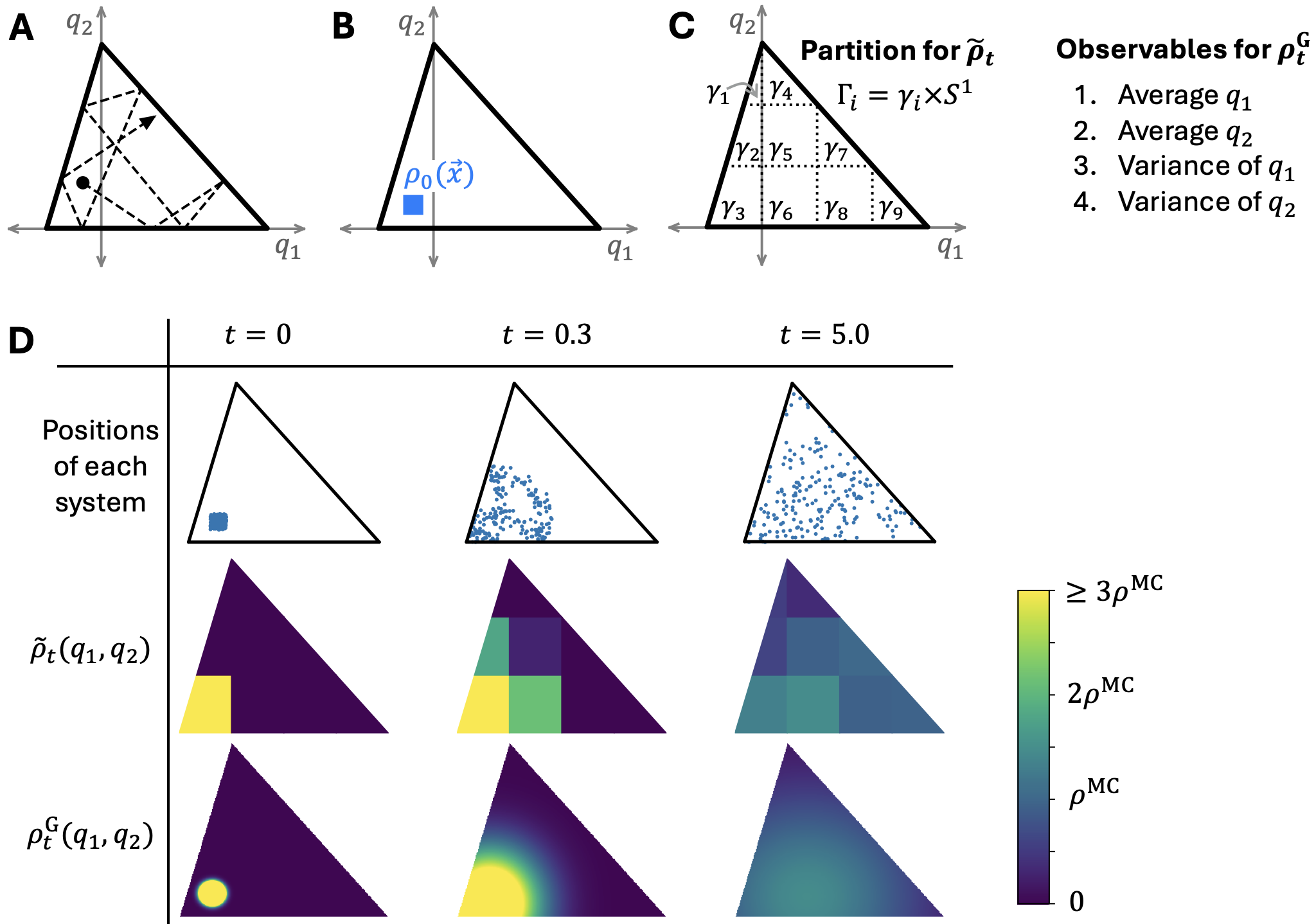}
    \caption{A particle in a 2D triangular box with vertices at (-0.3,0), (0.9,0), and (0,1). (A) Sample trajectory (dotted line). (B) Initial phase space distribution $\rho_0(\vec x)$ (blue rectangle). (C) Left: Partition of phase space used to compute $\tilde\rho_t(\vec x)$. Right: Observables used to compute $\rho^\text{G}_t(\vec x)$. (D) Simulation of an ensemble of 200 systems. Left, center, and right columns show the system at times $t=0$, 0.3, and 5.0. Top row: Positions of each system in the ensemble. Middle row: Standard coarse-grained distribution $\tilde\rho_t$. Bottom row: Generalized coarse-grained distribution $\rho^\text{G}_t$. } \label{triangle}
\end{figure}

Consider a single particle in a two-dimensional triangular box, as illustrated in Fig. \ref{triangle}A. The Hamiltonian is $H(q_1,q_2,p_1,p_2) = p_1^2 + p_2^2+ V(q_1,q_2)$ where $V(q_1,q_2)$ is zero inside the triangle and $\infty$ outside the triangle, so that the particle reflects off the walls specularly.
This is known as a \textit{triangular billiard} system, and there is strong numerical evidence that triangular billiards are strong mixing if each angle of the triangle is an irrational multiple of $\pi$ \cite{casati1999mixing, zahradova2022impact}.
Fig. \ref{triangle}A shows an example trajectory.

Now consider an ensemble of such systems initially confined to the small blue rectangle shown in Fig. \ref{triangle}B. We will approximate the infinite ensemble by a finite ensemble of 200 systems. For each of the 200 systems, the initial position is drawn randomly (uniformly) from within the small blue rectangle in Fig. \ref{triangle}B. For each system, the initial momentum is drawn randomly (uniformly) from the unit circle in momentum space (i.e. $(p_1,p_2)$ space), so that the energy of each system is $E=1$, and the initial momentum direction is random. Hence, all systems are on the hypersurface $\Sigma_{E=1}=T\times S^1$ where $T$ is the set of $(q_1,q_2)$ contained within the triangle, $S^1$ is the unit circle in $(p_1,p_2)$ space, and $\times$ denotes the Cartesian product.

We will compare a standard coarse-graining with a generalized coarse-graining. For the standard coarse-graining, $\Sigma_E$ is partitioned into nine cells as illustrated in Fig.~\ref{triangle}C. The dashed lines indicate the boundaries between the cells in $(q_1,q_2)$ space, and the partition is independent of momentum. To be precise, the cells are $\Gamma_i=\gamma_i\times S^1$, where $\gamma_i$ are the cells in $(q_1,q_2)$ space and $S^1$ is the unit circle in $(p_1,p_2)$ space.

For the generalized coarse-graining, suppose that the mean values and variances of particle position are observable. Hence, the observables are
\begin{align}
	a_1(q_1,q_2,p_1,p_2) &= q_1 \label{a1} \\
	a_2(q_1,q_2,p_1,p_2) &= q_2 \\
	a_3(q_1,q_2,p_1,p_2) &= q_1^{\,2} \\
    a_4(q_1,q_2,p_1,p_2) &= q_2^{\,2} \label{a4}
\end{align}
For this system with these observables, the integral in Eq.~\ref{g} can be efficiently solved numerically (python code is provided, see link below). This makes the optimization problem of minimizing $g_t(\lambda_0,...,\lambda_M)$ very fast with standard optimization packages.

Fig.~\ref{triangle}D (top row) shows the position of each of the 200 systems at three values of $t$, showing how the systems spread out over the triangle as $t$ increases. The middle row and bottom row show, respectively, the standard coarse-grained distribution $\tilde\rho_t(q_1,q_2)$ and the generalized coarse-grained distribution $\rho^\text{G}_t(q_1,q_2)$, at each time. Note that both $\tilde\rho_t$ and $\rho^\text{G}_t$ depend only on the position coordinates $(q_1,q_2)$ because both the partition and the observables $a_1,...,a_4$ are independent of momentum. Fig.~\ref{triangle}D illustrates how coarse-grained distributions that are initially non-uniform approach uniformity as $t\to\infty$

%\tilde\rho_t$ and $\rho^\text{G}_t$ are initially very far from uniform, and they approach the uniform microcanonical distribution as $t\to\infty$.
%Both the standard and generalized coarse-grained distributions illustrate how an initially far-from-equilibrium ensemble approaches the uniform microcanonical ensemble as $t$ increases.

\begin{figure}
    \centering
    \includegraphics[width=.98\linewidth]{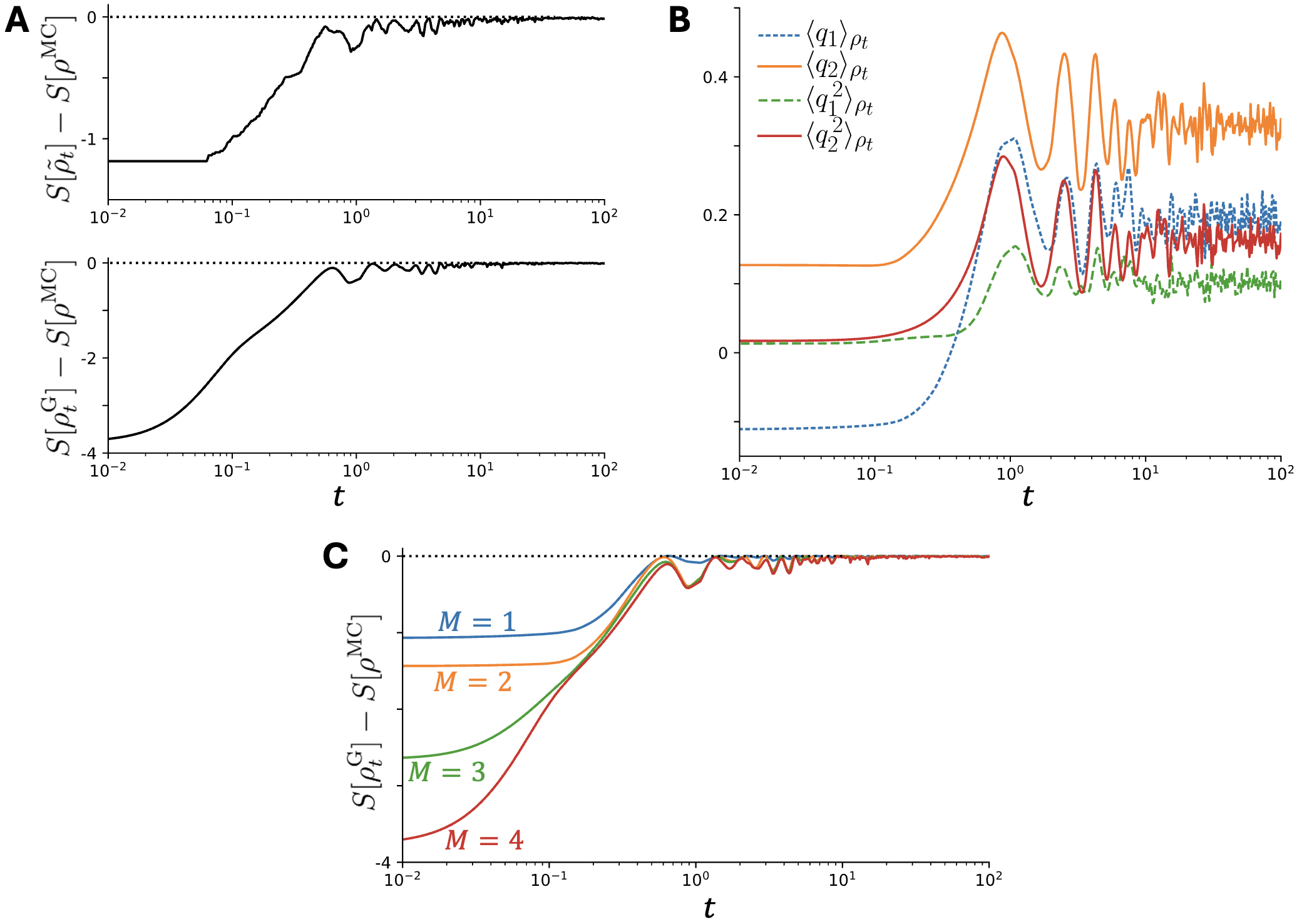}
    \caption{Entropy and ensemble averages of an ensemble of 200 particles in the triangular box. (A) Entropy vs time for the standard coarse-graining $\tilde \rho_t(\vec x)$ (top) and the generalized coarse-graining $\rho^\text{G}_t(\vec x)$ (bottom). A log-scale x-axis is used to make the dynamics at small $t$ more visible. (B) Ensemble averages $\langle q_1\rangle_{\rho_t}$, $\langle q_2\rangle_{\rho_t}$, $\langle q_1^{\,2}\rangle_{\rho_t}$, and $\langle q_2^{\,2}\rangle_{\rho_t}$ vs time. (C) Entropy of the generalized coarse-grained distribution using different numbers of observables, illustrating that including additional observables decreases entropy. Blue curve ($M=1$): $a_1$ is the only observable. Orange curve ($M=2$): $a_1$ and $a_2$ are observable. Green curve ($M=3$): $a_1$, $a_2$, and $a_3$ are observable. Red curve ($M=4$): $a_1$, $a_2$, $a_3$, and $a_4$ are observable.}
    \label{sims}
\end{figure}

Fig.~\ref{sims}A shows the entropy vs. $t$ for $\tilde\rho_t$ (top) and $\rho^\text{G}_t$ (bottom). $S[\tilde\rho_t]$ has a jagged appearance because it is a piecewise constant function of $t$, with discontinuity at any $t$ where a particle crosses from one cell to another. In contrast, $\rho^\text{G}_t(\vec x)$ is smooth because the observables $a_1$, $a_2$, $a_3$, and $a_4$ are smooth functions. As expected, both entropies approach the maximal value $S[\rho^\text{MC}]$. However, their approach to $S[\rho^\text{MC}]$ is not monotonic, and small fluctuations in entropy persist even as $t\to\infty$. These fluctuations decrease as the size of the ensemble grows, and they disappear in the limit of an infinite ensemble (as guaranteed by Theorem 2). However, the non-monotonicity of the entropy remains even in the limit of an infinite ensemble. Note also that $S[\tilde\rho_t]$ remains constant until $t>0.06$ because $S[\tilde\rho_t]$ is bounded from below: when all systems are within the same cell $i$, then $S[\tilde\rho_t]=k_B\log \mu(\Gamma_i)$. This contrasts with $S[\rho^\text{G}_t]$, which becomes arbitrarily negative if the variance in particle positions becomes arbitrarily small. Fig. S2 (Supplementary Material) plots the data of Fig.~\ref{sims}A top and bottom together on a log-log scale to aid visual comparison.

Fig.~\ref{sims}B shows the observable averages $\langle q_1\rangle_{\rho_t}$, $\langle q_2\rangle_{\rho_t}$, $\langle q_1^{\,2}\rangle_{\rho_t}$, and $\langle q_2^{\,2}\rangle_{\rho_t}$. At $t=0$, these are far from their microcanonical averages, and they approach and fluctuate around their microcanonical averages as $t\to\infty$. For an infinite ensemble, the ensemble averages would converge exactly to their microcanonical averages (with no fluctuations) due to weak convergence. For the finite ensemble of 200 systems, small fluctuations persist even as $t\to\infty$.
%, but the magnitude of these fluctuations decays as the number of systems in the ensemble increases.

Fig.~\ref{sims}C illustrates how adding additional observables decreases the coarse-grained entropy. The four curves, labelled $M=1$ through $M=4$, show the generalized coarse-grained entropy computed using 1 through 4 observables. Specifically, the blue curve ($M=1$) uses only $a_1$, the orange curve ($M=2$) uses $a_1$ and $a_2$, the green curve ($M=3$) uses $a_1$, $a_2$, and $a_3$, and the red curve ($M=4$) uses all four observables. This illustrates that entropy is an observer-dependent quantity that is lower for observers with more information. Nevertheless, all observers agree in the $t\to\infty$ limit.

\section{Discussion}

Briefly summarizing our results, we have proven a theorem (Theorem 1) showing that the time evolution of phase space distributions obeying Liouville's equation is an isometry according to a broad class of statistical distance metrics---specifically, metrics of the form of Eqs.~\ref{dist} and \ref{I}, which includes all the metrics listed in Table 1. This means that, according to these metrics, an ensemble never gets any ``closer" to an equilibrium ensemble. However, it is known that coarse-grained distributions do converge to $\rho^\text{MC}$ for strong mixing systems \cite{dougherty1994foundations,mackey1989dynamic,kozlov2007fine, safranek2020classical}. Our second contribution is to generalize the definition of coarse-graining to remove the need for an arbitrary partition of phase space.
%This puts generalized coarse-graining on the same footing as weak convergence; both are defined in terms of observables, which can be any integrable function on phase space.
For strong mixing systems, the generalized coarse-grained distribution $\rho^\text{G}_t(\vec x)$ converges pointwise to $\rho^\text{MC}$ for any list of observables (Theorem 2). Likewise, the entropy of $\rho^\text{G}_t(\vec x)$ increases to its maximal value.

Let us consider the physical scenarios described by distributions $\rho_t(\vec x)$ obeying Liouville's equation. These are scenarios in which an initial condition $\vec x_0$ is drawn at random from some non-equilibrium distribution $\rho_0(\vec x)$, and one is interested in the probability distribution of the state $\vec x(t)$ at later times. Weak convergence to $\rho^\text{MC}$ implies that, in the $t\to\infty$ limit, the probability that $\vec x(t)$ is in some region $B\subseteq\Sigma_E$ is simply the volume of $B$ times the microcanonical probability density, namely $\mu(B)\rho^\text{MC}=\mu(B)/\mu(\Sigma_E)$ (see sections \ref{mixing} and \ref{weak_and_strong}). In other words, the system ``forgets" its initial condition.
%Strong mixing and weak convergence to $\rho^\text{MC}$ imply that, as $t\to\infty$, the system ``forgets" its initial condition so that the probability that $\vec x(t)$ is in any region $B$ is simply proportional to the volume of $B$, namely $\mu(B)/\mu(\Sigma_E)$. 
Weak convergence can be of practical utility in computational studies where one wants to sample from the microcanonical distribution. If one initializes the system by randomly drawing $\vec x(0)$ from any non-equilibrium distribution, then simulates the dynamics for a sufficiently long time $t$, the resulting state $\vec x(t)$ will be distributed microcanonically. Using this method to compute microcanonical averages by averaging over repeated simulations can be faster than estimating microcanonical averages by computing a time average along a single long-time trajectory \cite{calvo2002sampling}. The required simulation time is given by the timescale on which $S[\rho^\text{G}_t]$ approaches $S[\rho^\text{MC}]$ (which depends upon the observables used to compute $\rho^\text{G}_t$). In experiments on isolated systems, weak convergence implies that averages over repeated trials will converge to the microcanonical average, if each trial is initialized with approximately the same energy. The details of the experimental setup will dictate the initial distribution $\rho_0(\vec x)$ according to which the initial state $\vec x(0)$ is randomly initialized, and weak convergence implies that $\vec x(t)$ forgets this initial state and becomes microcanonically distributed for large $t$.

A central question in the foundations of statistical mechanics is: Why do individual isolated macroscopic systems exhibit equilibrium behavior, where a system's macroscopic properties settle to nearly constant values that undergo only exceedingly rare fluctuations? Weak convergence alone does not explain this, as weak convergence is a statement about ensemble averages rather than individual systems. Rather, it is an aspect of the thermodynamic limit---namely, that macroscopic observable functions approach uniformity on $\Sigma_E$ in the limit of large particle number \cite{lanford1973entropy, ruelle1965correlation, lof1979,khinchin1949mathematical} (see also \cite{uffink2006compendium,tumulka2019})---that explains why macroscopic properties of individual systems settle to nearly constant values. This feature of the thermodynamic limit implies that macroscopic systems will exhibit equilibrium behavior after traversing only a small portion of $\Sigma_E$ \cite{goldstein2001boltzmann, bricmont2022making, werndl2023does,albert2003time}. Importantly, this can occur without strong mixing or ergodicity \cite{goldstein2001boltzmann, bricmont2001bayes,khinchin1949mathematical,bricmont2022making}. In summary, while the weak convergence of $\rho_t(\vec x)$ to $\rho^\text{MC}$ is indeed relevant to particular computations and experiments, it misses a key aspect of (and is not necessary for) the thermalization of macroscopic systems.

%A central question in the foundations of statistical mechanics is: Why do \textit{individual} isolated macroscopic systems exhibit equilibrium behavior, where a system's macroscopic properties settle to nearly constant values that undergo only exceedingly rare fluctuations? Weak convergence does not imply anything about the behavior of individual systems, only about ensemble averages. Rather, it is an aspect of the thermodynamic limit---namely, that macroscopic observable functions approach uniformity on $\Sigma_E$ in the limit of large particle number \cite{lanford1973entropy,ruelle1965correlation,lof1979,khinchin1949mathematical} (see also \cite{uffink2006compendium,tumulka2019})---that explains why macroscopic properties of individual systems approach equilibrium values. This feature of the thermodynamic limit gives rise to \textit{typicality}: the phenomenon that nearly all microstates have nearly the same, `typical', macroscopic properties \cite{goldstein2001boltzmann}. Due to typicality, an individual trajectory initialized with non-equilibrium properties will typically approach equilibrium after traversing only a small portion of $\Sigma_E$ \cite{goldstein2001boltzmann,werndl2023does,albert2003time}; this is true regardless of strong mixing or any other ergodic properties \cite{goldstein2001boltzmann,lanford1973entropy}. In summary, the weak convergence of $\rho_t(\vec x)$ to $\rho^\text{MC}$ is indeed relevant to particular computations and experiments, but it does not provide a suitable description of the thermalization of macroscopic systems.

\section*{Acknowledgements}

I am grateful to Robijn Bruinsma, Giovanni Zocchi, Jacob Pierce, Tyler Carbin, and the participants of the UCLA Soft Condensed Matter Journal Club, for helpful discussions and comments. This work was supported by the NSF Graduate Research Fellowship Program (NSF Grant No. DGE-2034835).

\section*{Code availability}
Python code to reproduce all results and figures is available at:\\ https://github.com/cbarkan1/convergence-to-microcanonical

\appendix

\section{Proof of Theorem 1}

We prove two theorems in this section. The first (Theorem 1A) is a simplified version of Theorem 1 in which $\rho_t(\vec x)$ is assumed to be differentiable on the full phase space $\mathcal{P}$, and $F(f,g)$ and $d(I)$ (in Eqs.~\ref{dist} and \ref{I}) are also assumed to be differentiable. These assumptions allow for a simple and transparent proof, but they rule out distributions with support on a constant energy hypersurface. Then, we prove Theorem 1, which does not require these assumptions, but which has a more complicated proof.

\subsection{Theorem 1A}

\textbf{Theorem 1A:} Let $\rho_t(\vec x)$ and $\eta_t(\vec x)$ be differentiable ($C^1$) functions on phase space $\mathcal{P}$ that obey Liouville's equation. Define $\text{dist}[\rho_t(\vec x),\eta_t(\vec x)]$ according to Eqs.~\ref{dist} and \ref{I} where $d:\mathbb{R}\to\mathbb{R}$ and $F:\mathbb{R}\to\mathbb{R}_{\geq0}$ are differentiable ($C^1$). Then, $\frac{\text d}{\text d t}\text{dist}[\rho_t(\vec x),\eta_t(\vec x)]=0$; in other words, the distance is unchanging in time.

\medskip
\noindent\textbf{Proof: } Define $\mathcal{F}_t(\vec x)=F(\rho_t(\vec x),\eta_t(\vec x))-F(0,0)$. Then 
\begin{align}
    \partial_t\mathcal{F}_t(\vec x) &= (\partial_1F)\partial_t\rho_t(\vec x) + (\partial_2F)\partial_t\eta_t(\vec x) \\
    &= (\partial_1F)(-\vec v(\vec x)\cdot\nabla\rho_t) + (\partial_2F)(-\vec v(\vec x)\cdot\nabla\eta_t)
\end{align}
where we have used Eq.~\ref{Liouville}, and $\vec v(\vec x)$ is as defined below Eq.~\ref{Liouville}. Simplifying, we obtain

\begin{align}
    \partial_t\mathcal{F}_t(\vec x) &= -\vec v(\vec x) \cdot ( (\partial_1F)\nabla\rho_t + (\partial_2F)\nabla\eta_t) \\
    &= -\vec v(\vec x) \cdot \nabla \mathcal{F}_t(\vec x)
\end{align}
Therefore, $\mathcal{F}_t(\vec x)$ itself obeys Eq.~\ref{Liouville}. Hence, $\frac{\text d}{\text d t}\int_\mathcal{P}\mathcal{F}_t(\vec x)d\mu=0$ (this is because Eq.~\ref{Liouville} is a continuity equation that conserves total probability). From Eq.~\ref{I}, $\int_\mathcal{P}\mathcal{F}_t(\vec x)d\mu=I[\rho_t(\vec x),\eta_t(\vec x)]$, therefore,
$\frac{\text d}{\text d t}\text{dist}[\rho_t(\vec x),\eta_t(\vec x)] = d'(\int_\mathcal{P}\mathcal{F}_t(\vec x)d\mu)\frac{\text d}{\text d t}\int_\mathcal{P}\mathcal{F}_t(\vec x)d\mu=0$. $\square$

\subsection{Theorem 1}

To prove Theorem 1, we will first prove the following lemma, which is equivalent to Theorem 1 except that it assumes that $F(f,g)$ is a simple function (i.e. a sum of indicator functions). Then, we use the lemma to prove Theorem 1.

Define the functional $I_s[f(\vec x),g(\vec x)]$ according to Eq.~\ref{Is}. $I_s[f(\vec x),g(\vec x)]$ is equivalent to the functional $I[f(\vec x),g(\vec x)]$ defined in Eq.~\ref{I}, except that it is defined in terms of a simple function; the subscript $s$ stands for ``simple function".
\begin{equation}\label{Is}
	I_s[f(\vec x),g(\vec x)] = \int_P F_s(f(\vec x),g(\vec x)) d\mu
\end{equation}
%where $p$ is either $P$ or a constant-energy hypersurface $\Sigma_E$
where $F_s:\mathbb{R}^2\to\mathbb{R}$ is any simple function,
\begin{equation}
    F_s(y_1,y_2)=\sum_{m=1}^M f_m 1_{T_m}(y_1,y_2)
\end{equation}
where the sets $T_m$ are measurable subsets of $\mathbb{R}^2$ and $1_{T_m}(y_1,y_2)$ denotes the indicator function of $T_m$, defined as $1_{T_m}(y_1,y_2)=1$ if $(y_1,y_2)\in T_m$, and $1_{T_m}(y_1,y_2)=0$ otherwise.

\medskip
\noindent\textbf{Lemma 1: } If $\rho_t(\vec x)$ and $\eta_t(\vec x)$ obey Liouville's equation, then $I_s[\rho_t(\vec x),\eta_t(\vec x)]$ is independent of $t$.

\medskip
\noindent\textbf{Proof: } Define $\omega_m(t) = \{\vec x\in P : (\rho_t(\vec x),\eta_t(\vec x))\in T_m \}$. Then,
\begin{align}
    \omega_m(t) &= \{ \vec x : (\rho_0(\phi_{-t}\vec x),\eta_0(\phi_{-t}\vec x))\in T_m \}\\
    &= \{ \phi_{t}\vec x : (\rho_0(\vec x),\eta_0(\vec x))\in T_m \}\\
    &= \phi_t\omega_m(0)
\end{align}
Therefore,
\begin{align}
   I_s[\rho_t(\vec x),\eta_t(\vec x)] &= \int_P F_s(\rho_t(\vec x),\eta_t(\vec x)) d\mu \\
    &= \sum_{m=1}^M f_m \int_P 1_{T_m}(\rho_t(\vec x),\eta_t(\vec x))d\mu \\
    &= \sum_{m=1}^M f_m \int_{\omega_m(t)}d\mu = \sum_{m=1}^M f_m \mu(\omega_m(t)) \\
    &= \sum_{m=1}^M f_m \mu(\phi_t\omega_m(0))\label{mu_t} \\
    &= \sum_{m=1}^M f_m \mu(\omega_m(0))\label{mu_0}
\end{align}
where in going from Eq.~\ref{mu_t} to \ref{mu_0} we have used the fact that $\mu$ is an invariant measure, i.e. $\mu(\phi_t A)=\mu(A)$ for any measurable $A$. Eq. \ref{mu_0} is manifestly independent of $t$. $\square$

\medskip
\noindent\textbf{Theorem 1: } Let $\rho_t(\vec x)$ and $\eta_t(\vec x)$ be two phase space distributions obeying Liouville's equation and define $\text{dist}[\rho_t(\vec x),\eta_t(\vec x)]$ according to Eqs.~\ref{dist} and \ref{I} for any $d:\mathbb{R}\to\mathbb{R}$, any Lebesgue integrable $F:\mathbb{R}\to\mathbb{R}_{\geq0}$, and any invariant $P\subseteq\mathcal{P}$ with $d\mu$ the corresponding invariant measure. Then, $\text{dist}[\rho_t(\vec x),\eta_t(\vec x)]$ is independent of time $t$.

\medskip
\noindent\textbf{Proof: } Consider a sequence of simple functions $F^{(M)}:\mathbb{R}^2\to\mathbb{R}$ such that $F^{(M)}\to F$ for $M\to\infty$ and $F^{(M)}(\vec y) \leq F(\vec y)$ for all $\vec y\in\mathbb{R}^2$. The existence of such a sequence $F^{(M)}$ is a standard result (see Theorem 11.20 in \cite{rudin1976principles}). These functions are of the form $F^{(M)}(y_1,y_2)=\sum_{m=1}^M f_m^{(M)}1_{T^{(M)}_m}(y_1,y_2)$ where, as before, $T^{(M)}_m$ are measurable subsets of $\mathbb{R}^2$ and $1_{T^{(M)}_m}(y_1,y_2)$ is the characteristic function of $T^{(M)}_m$.

By Lebesgue's Monotone Convergence Theorem \cite{rudin1976principles}, 
\begin{equation}
    \lim_{M\to\infty}\int_P F^{(M)} d\mu = \int_P F d\mu = I[\rho_t(\vec x),\eta_t(\vec x)]
\end{equation}
where the second step is merely the definition of $I$ given in Eq.~\ref{I}. Also, from Lemma 1 we know
\begin{equation}
    \int_P F^{(M)}d\mu = \sum_{m=1}^M f^{(M)}_m \mu(\omega^{(M)}_m(0))
\end{equation}
where, as before, $\omega^{(M)}_m(t)= \{\vec x\in P : (\rho_t(\vec x),\eta_t(\vec x))\in T^{(M)}_m \}$. Hence, we have
\begin{equation}
    I[\rho_t(\vec x),\eta_t(\vec x)] = \lim_{M\to\infty} \sum_{m=1}^M f^{(M)}_m \mu(\omega^{(M)}_m(0))
\end{equation}
which is manifestly independent of $t$. Therefore,
\begin{equation}
    \text{dist}[\rho_t(\vec x),\eta_t(\vec x)] = d(I[\rho_t(\vec x),\eta_t(\vec x)])
\end{equation}
is also independent of $t$. $\square$

\section{Comparing $\bm{\rho^\text{G}_t(\vec x)}$ to E.T. Jaynes' maximum entropy distribution}

Although $\rho^\text{G}_t(\vec x)$ is similar to E.T. Jaynes' maximum entropy distribution \cite{jaynes1957information} in that both are defined in terms of a maximum entropy optimization problem, there is a key difference: $\rho^\text{G}_t(\vec x)$ is guaranteed to converge to $\rho^\text{MC}$ for strong mixing systems, whereas Jaynes' maximum entropy distribution does not converge to either the canonical nor microcanonical distribution, except under a restrictive assumption. This is explained below.

E.T. Jaynes introduced the maximum entropy distribution $\rho^\text{J}_t(\vec x):\mathcal{P}\to\mathbb{R}_{\geq 0}$ defined as
\begin{equation}\label{Jaynes_def}
\begin{split}
	\rho^\text{J}_t(\vec x) = \argmax_{\eta(\vec x)\in L^2(\mathcal{P})}\; \left[-\int_\mathcal{P} \eta(\vec x)\log \eta(\vec x)d\vec x \right] \quad \text{subject to}& \quad \int_{\mathcal{P}} \eta(\vec x)d\vec x = 1 \\
	\text{and}& \quad \langle a_i \rangle_{\eta} = \langle a_i \rangle_{\rho_t} \text{ for } i=1,...,M
\end{split}
\end{equation}
where, unlike in Eq.~\ref{rhoG}, $\langle a_i \rangle_{\eta}=\int_\mathcal{P} a_i(\vec x)\eta(\vec x)d\vec x$ and $\langle a_i \rangle_{\rho_t}=\int_\mathcal{P} a_i(\vec x)\rho_t(\vec x)d\vec x$, and $L^2(\mathcal{P})$ denotes the space of square integrable functions on $\mathcal{P}$. The solution to this optimization problem is
\begin{equation}\label{Jaynes}
	\rho^\text{J}_t(\vec x) = \exp\left(-1-\mu_{0}^*(t) - \sum_{i=1}^M \mu_{i}^*(t) a_i(\vec x)\right)
\end{equation}
where $\mu_{i}^*(t)$ are chosen to solve the equations $\langle a_i \rangle_{\rho^\text{J}_t} = \langle a_i \rangle_{\rho_t}$.

The difference between $\rho^\text{G}_t(\vec x)$ and $\rho^\text{J}_t(\vec x)$ is that $\rho^\text{G}_t(\vec x)$ is defined on a hypersurface $\Sigma_E$ (and all integrals in the optimization are taken over $\Sigma_E$), whereas $\rho^\text{J}_t(\vec x)$ is defined on the full phase space $\mathcal{P}$. 

%The fact that $\rho^\text{J}_t(\vec x)$ is defined on the full phase space means that it does not, in general, dynamically evolve into an equilibrium distribution under the dynamics of Liouville's equation, as explained below.

Under the dynamics of Liouville's equation, $\rho^\text{J}_t(\vec x)$ will not converge to the canonical distribution $\rho^\text{can}(\vec x)$ unless either (i) $H^n(\vec x)$ is \textit{not} observable for any $n>1$, meaning that the variance in energy cannot be observed, or (ii) $\langle H^n(\vec x)\rangle_{\rho_t}$ is \textit{already} equal to its canonical value $\langle H^n(\vec x)\rangle_{\rho^\text{can}}$ at $t=0$. This condition is due to the fact that $\langle H^n(\vec x)\rangle_{\rho_t}$ is a constant of the motion, so its value must either be at equilibrium initially (in which case the approach to equilibrium is not explained), or the experimenter must be ignorant of its value, in order for $\rho^\text{J}_t(\vec x)$ to approach $\rho^\text{can}(\vec x)$.

Similarly, $\rho^\text{J}_t(\vec x)$ does not evolve to the microcanonical distribution except in the special case that the variance in energy \textit{is} an observable \textit{and} is arbitrarily small. In the limit that $\langle H^2(\vec x)\rangle_{\rho_t} - \langle H(\vec x)\rangle_{\rho_t}^2\to 0$, $\rho^\text{J}_t(\vec x)$ becomes zero everywhere except on a hypersurface $\Sigma_E$, and $\rho^\text{J}_t(\vec x)\to\rho^\text{G}_t(\vec x)$.

%In conclusion, our generalized coarse-grained distribution dynamically evolves to microcanonical equilibrium for strong mixing systems obeying Liouville's equation. However, E.T. Jaynes' maximum entropy distribution typically does not evolve to canonical nor microcanonical equilibrium under Liouville's equation.

\section{Solving the optimization problem for $\bm\rho^\text{G}_{\bm t}$}
The first step for computing $\rho^\text{G}_t$ is to write the Lagrangian for the optimization problem. Note that this is the Lagrangian in the context of optimization theory \cite{boyd2004convex}, which is the entropy functional being maximized plus terms with Lagrange multiplies for each constraint. It is not the physical Lagrangian associated with the Hamiltonian $H(\vec x)$. The Lagrangian is
\begin{equation}
	L_t[\eta(\vec x),\lambda_0,...,\lambda_M] = S[\eta(\vec x)] - \lambda_0\left( \int_{\Sigma_E} \eta(\vec x)d\mu-1 \right) - \sum_{i=1}^M \lambda_i \left( \int_{\Sigma_E} \eta(\vec x)a_i(\vec x)d\mu - \langle a_i\rangle_{\rho_t} \right)
\end{equation}
where $M$ is the number of observables available to the experimenter. The subscript $t$ in $L_t[\eta(\vec x),\lambda_1,...,\lambda_M]$ indicates that the Lagrangian carries a time dependence due to the time dependence of $\langle a_i\rangle_{\rho_t}$. The variables $\lambda_i \in\mathbb{R}$ are the Lagrange multipliers for the problem. $\lambda_0$ corresponds to the normalization constraint, and $\lambda_i$ for $i=1,...,M$ correspond to the constraints on the ensemble averages. Importantly, the values $\langle a_i\rangle_{\rho_t}$ are considered to be known to the experimenter, so they are treated as known constants in the optimization problem.

The concavity of $S[\eta(\vec x)]$ guarantees that $L_t$ has a single stationary point corresponding to the unique optimum \cite{boyd2004convex}. The stationarity condition $\delta L/\delta\eta=0$ is solved by a function $\eta^*(\vec x,\lambda_0,...,\lambda_M)$ that maximizes $L_t$ under fixed $\lambda_i$. To simplify notation, we denote the list of Lagrange multipliers as $\bm\lambda =(\lambda_0,...,\lambda_M)$. Then, the function $\eta^*(\vec x,\bm\lambda)$ that solves $\delta L/\delta\eta=0$ is
\begin{equation}
	\eta^*(\vec x;\bm\lambda) = \exp\left(-1-\lambda_0 - \sum_{i=1}^M \lambda_i a_i(\vec x)\right)
\end{equation}

We now use the method of Lagrange duality to determine the Lagrange multipliers $\bm\lambda$. An important result in the theory of convex optimization says that the Lagrange dual function $g_t(\bm\lambda)=\max_{\eta(\vec x)} L_t[\eta(\vec x),\bm\lambda]$ is a convex function with a unique minimum (because $S[\eta(\vec x)]$ is concave), and the $\bm\lambda^*(t)$ that minimize $g_t(\bm\lambda)$ are the Lagrange multipliers that solve the original optimization problem (when the duality gap is zero, as is the case here) \cite{boyd2004convex}. In other words,
\begin{equation}\label{rhoG_A}
	\rho_t^\text{G}(\vec x) = \eta^*(\vec x; \bm\lambda^*(t))
\end{equation}
where $\bm\lambda^*(t)$ is the unique minimum of the function $g_t(\bm\lambda)$ defined as
\begin{align}
	g_t(\bm\lambda) &= \max_{\eta(\vec x)\in L^2} L_t[\eta(\vec x),\bm\lambda] = L_t[\eta^*(\vec x;\bm\lambda),\bm\lambda] \\
	&= \int_{\Sigma_E} \exp\left( -1-\lambda_0-\sum_{i=1}^M\lambda_ia_i(\vec x) \right)d\mu + \lambda_0 + \sum_{i=1}^M \lambda_i\langle a_i\rangle_{\rho_t}\label{g_A}
\end{align}
where $L^2$ is the space of square integrable functions on $\Sigma_E$. Eqs.~\ref{rhoG_A} and \ref{g_A} are equivalent to Eqs.~\ref{rhoG_solved} and \ref{g}, completing the derivation.

%To be clear, $t$ is a fixed constant in the optimization, because computing $\rho^\text{G}_t$ requires solving the optimization at the fixed time $t$.

%The original infinite dimensional constrained optimization problem has now been reduced to an $N$ dimensional unconstrained optimization problem. The convexity of $g_t(\bm\lambda)$ guarantees that a gradient descent algorithm will converge to the unique optimum $\bm\lambda^*_t$. In simple cases, the integral in Eq.~\ref{g_phys} can be solved analytically, which allows very fast computation of $\bm\lambda^*_t$, and thus of $\rho^\text{G}_t(\vec x)=\rho^*(\vec x;\bm\lambda^*_t)$, with standard optimization computer packages. This is done for an example in the following section.

\subsection{The equivalence of Eq.~\ref{maxS} and Eq.~\ref{SCG}}

Here, we show that Eq.~\ref{maxS} is equivalent to Eq.~\ref{SCG}. We start with $\tilde\rho_t(\vec x)$ as defined by Eq.~\ref{maxS}, which is the special case of $\rho^\text{G}_t(\vec x)$ in which $a_i(\vec x)=1_{\Gamma_i}(\vec x)$, where $\Gamma_i$ are sets that form a partition of $\Sigma_E$. By solving the optimization problem using Eqs.~\ref{rhoG} and \ref{g}, we show that the solution to Eq.~\ref{maxS} is Eq.~\ref{SCG}. 

Suppose that $a_i(\vec x)=1_{\Gamma_i}(\vec x)$, where $\Gamma_i$ are sets that form a partition of $\Sigma_E$. Then, Eq.~\ref{rhoG_solved} reduces to
\begin{equation}\label{rhoG_S}
    \rho^\text{G}_t(\vec x) = \tilde\rho_t(\vec x) = \exp(-1-\lambda_0-\lambda_{i(\vec x)})
\end{equation}
where $i(\vec x)$ is the index of the cell containing point $\vec x$, i.e. $\vec x\in\Gamma_{i(\vec x)}$. Eq.~\ref{g} reduces to
\begin{equation}\label{g_S}
    g_t(\lambda_0,...,\lambda_M) = e^{-1-\lambda_0}\sum_{i=1}^M e^{- \lambda_i}\mu(\Gamma_i) + \lambda_0 + \sum_{i=1}^M \lambda_i\langle 1_{\Gamma_i}\rangle_{\rho_t}
\end{equation}
The first order condition for minimizing Eq.~\ref{g_S}, $\partial g_t/\partial\lambda_i=0$, gives
\begin{equation}
    \lambda_i = \log \frac{\mu(\Gamma_i)}{\langle1_{\Gamma_i}\rangle_{\rho_t} e^{1+\lambda_0}}
\end{equation}
for $i=1,...,M$. Inserting this into Eq.~\ref{rhoG_S} gives
\begin{equation}
    \tilde\rho_t(\vec x) = \frac{\langle1_{\Gamma_{i(\vec x)}}\rangle_{\rho_t}}{\mu(\Gamma_{i(\vec x)})} = \frac{1}{\mu(\Gamma_{i(\vec x)})}\int_{\Gamma_{i(\vec x)}}\rho_t(\vec x')d\mu
\end{equation}
which is Eq.~\ref{SCG}.

\section{Proof of Theorem 2}

\noindent\textbf{Theorem 2: }Let $\rho_t(\vec x)$ be a normalized and square integrable distribution on $\Sigma_E$ obeying Liouville's equation for a strong mixing system. Define $\rho^\text{G}_t(\vec x)$ according to Eq.~\ref{rhoG}. Then,
\begin{enumerate}
	\item $\lim_{t\to\infty}\rho^\text{G}_t(\vec x)=\rho^\text{MC}$ for all $\vec x$, i.e. pointwise convergence.
	\item $\lim_{t\to\infty}S[\rho^\text{G}_t(\vec x)] = S[\rho^\text{MC}]$.
\end{enumerate}

\medskip
\noindent\textbf{Proof: } Note the following three facts:
\begin{enumerate}[label=(\alph*)]
    \item If $\langle a_i\rangle_{\rho_t}=\langle a_i\rangle_{\rho^\text{MC}}$ for $i=1,...,M$, then $\rho^\text{G}_t=\rho^\text{MC}$. This is because $\rho^\text{MC}$ solves
\begin{equation}
	\rho^\text{MC} = \argmax_{\eta(\vec x)\in L^2}\; S[\eta(\vec x)] \quad  \text{subject to} \quad  \int_{\Sigma_E} \eta(\vec x)d\mu = 1
\end{equation}
and $\rho^\text{MC}$ satisfies the additional constraints of Eq.~\ref{rhoG} when $\langle a_i\rangle_{\rho_t}=\langle a_i\rangle_{\rho^\text{MC}}$.

    \item $\lim_{t\to\infty}\langle a_i\rangle_{\rho_t}=\langle a_i\rangle_{\rho^\text{MC}}$ for strong mixing dynamics due to weak convergence.
    \item The solution to the optimization problem of Eq.~\ref{rhoG} is continuous with respect to $\langle a_i\rangle_{\rho_t}$.
\end{enumerate}
Fact (c), together with (a) and (b), imply that $\lim_{t\to\infty}\rho^\text{G}_t=\rho^\text{MC}$, which proves (i). (ii) follows from (i) due to the continuity of $S[\rho(\vec x)]$. $\square$

\section*{References}
\bibliographystyle{unsrt}
\bibliography{main}

\end{document}